# Selecting cells in a raster database for maximal impact intervention in the presence of spatial interaction: Computational complexity of a Multiple vs. a Single Flow Direction Method

Grethell Castillo-Reyes[a,b], René Estrella[c], Karen Gabriels[c], Jos Van Orshoven[c], Floris Abrams[c] and Dirk Roose[a]

[a]*Department of Computer Science, KU Leuven, Belgium*
[b]*Data Representation and Analysis Center, University of Informatic Sciences, Cuba*
[c]*Division of Forest Nature and Landscape, Department of Earth and Environmental Sciences, KU Leuven, Belgium*

**ABSTRACT**

To minimize the sediment flowing to the outlet of a river catchment with minimal effort or cost, it is important to select the best areas to perform a certain intervention, e.g., afforestation. CAMF (Cellular Automata based heuristic for Minimizing Flow) is a method that performs this selection process iteratively in a raster geodatabase environment. To simulate the flow paths, the original CAMF uses a Single Flow Direction (SFD) algorithm. However, SFD fails to reflect the real nature of flow transport, especially in areas with low relief. This paper describes and analyzes the integration of a Multiple Flow Direction (MFD) algorithm in CAMF, in order to provide a more realistic flow simulation. We compare the computational complexity of CAMF-SFD and CAMF-MFD and we discuss the scalability w.r.t. the number of cells involved. We evaluate the behavior of both variants for sediment yield minimization by afforestation in two catchments with different properties.

## CRediT authorship contribution statement

**Grethell Castillo-Reyes:** Study, Conceptualization, Design, Software development, Analysis and interpretation of results, Writing. **René Estrella:** Generation of data, Analysis and interpretation of results, Review. **Karen Gabriels:** Generation of data. **Jos Van Orshoven:** Conceptualization, Supervision, Analysis and interpretation of results. **Floris Abrams:** Generation of data, Review. **Dirk Roose:** Conceptualization, Design, Methodology, Supervision, Analysis and interpretation of results, Review, Writing.

## 1. Introduction

Among the main issues analyzed in Spatial Decision Support Systems, land use planning is of vital importance. It comprises the management and modification of the natural environment, focusing on developing effective strategies for land conservation and urbanization (Randolph, 2012). It allows to select potential sites to carry out a certain action, taking into account criteria related to the use of the land depending on the application.

Sediment delivery to rivers reduces channel and reservoir capacity and also leads to water quality problems and biodiversity decline because of suspended mineral and organic substances (Drzewiecki and Mularz, 2008). These issues are particularly undesirable in regions where the river is used for drinking or irrigation water provision or for electricity production.

ORCID(s): 0000-0002-6922-7735 (G. Castillo-Reyes)

Selecting cells in a raster database for maximal impact intervention

Afforestation has proven to be effective to reduce sediment production and delivery and to alleviate the associated water quality problem (Costin, 1980; Nearing et al., 2005; Heil et al., 2007; Vanwalleghem, 2017). It is known to decrease runoff and protect the soil surface against the ability of raindrops and runoff to detach and transport sediments (B.-M. Vought et al., 1995; Piégay et al., 2013). A main question at the beginning of any afforestation project is to identify the most suitable sites to plant the trees. The discrimination between suitable and unsuitable areas for afforestation typically depends on several criteria adopted by forest planners, which can range from on-site and off-site environmental concerns to maximizing financial profits.

This spatial optimization problem can be formulated as a mathematical programming model that can be solved exactly, but typically requires a high amount of computational resources (Fischer and Church, 2003). While this approach has been used for many years in e.g., forest planning (Williams and Revelle, 1997), the size of the problems that can be handled in a practical way remains limited (Vanegas et al., 2011). Therefore, as an alternative, Vanegas (2010); Vanegas et al. (2012, 2014) proposed a Cellular Automata based heuristic for Minimizing Flow (CAMF) to locate the sites within a river catchment that should be afforested in order to minimize the amount of sediment that reaches the outlet. Although finding the optimal solution cannot be guaranteed in general for an heuristic method, they proved theoretically that CAMF obtains the optimal solution for a small raster under certain conditions. To simulate sediment transport CAMF uses a simple spatially distributed model, which is computationally efficient in order to be used in an optimization context.

A key issue in sediment flow simulation with CAMF is the spatial interaction (SI) among cells in the raster datasets, representing the catchment. SI refers to the fact that changes in the state of a location can have an impact on the state of neighboring or even distant locations (Gersmehl, 1970; Wang, 2017). In the case of CAMF, SI refers to the phenomenon that afforestation of a cell leads to changes of its characteristics that in turn affect the amount of sediment flowing from that cell into its downstream neighbor cells and eventually the sediment yield of the catchment.

The original implementation of CAMF uses a Single Flow Direction (SFD) model, which assumes that flow leaving a cell is delivered entirely to the neighbor at the lowest altitude. Despite the advantage of simplicity of SFD models, it has been suggested that they fail to reflect the real nature of surface transport processes (Quinn et al., 1991; Wilson and Gallant, 2000). Quinn et al. (1991) proposed a Multiple Flow Direction (MFD) model, which are based on the assumption that flow is distributed to one or more of the neighboring downslope cells. This model can represent sediment flow pathways more appropriately than SFD. With SFD even a tiny elevation difference between two neighboring cells can have a large effect, since it might determine which cell receives all outgoing flow. Comparatively, small elevation differences have a less important effect in an MFD algorithm, since cells with slight differences in elevation receive about the same amount of flow.

In this paper, we develop and analyze the MFD variant of CAMF (CAMF-MFD) and we show why the compu-tational cost substantially increases. Several experiments were carried out using different data-sets, with the aim of comparing the behaviour and performance of the two variants of CAMF.

This paper is organized as follows. Section 2 provides a general description of CAMF, the calculation of the sediment accumulation matrix and the main steps followed by CAMF to select the best areas for afforestation. It also describes particular characteristics of CAMF-SFD and its reformulation to use a MFD method. Section 3 presents the case studies used for experiments. In Section 4 the results of experiments are discussed and Section 5 presents some conclusions.

## 2. CAMF method for minimizing sediment flow

Using a rasterized database representing a river catchment, the CAMF method (Vanegas, 2010; Vanegas et al., 2012, 2014) selects cells, such that performing a certain action in these cells leads to the maximum reduction of the sediment yield, i.e. the amount of sediment in the outlet, denoted by *SY*. The cells are selected from a list of candidate cells, since the action can be performed only in cells corresponding to some land use types. Because of SI, cells are selected iteratively. In each iteration, CAMF computes for every candidate cell the sediment yield reduction, denoted by *SYR*, that will be achieved if the action is performed in that cell. The candidate cells are ranked based on their *SYR* values. The action is performed in the cell(s) with the maximum *SYR*. Although the core structure of CAMF can be applied for different sorts of flow, until now CAMF has been used exclusively for afforestation planning. Therefore we will now describe CAMF in the context of afforestation.

The input data required by CAMF consist of several raster data-sets, each of them representing one characteristic of the area under study:

- Digital Elevation Model (DEM) to compute the gradients and the flow direction matrix to determine the flow paths that sediment follows within a catchment.

- Land-use types map or coverage map to identify the cells on which action can be taken.

- Initial local sediment production ($ton\ ha^{-1}\ yr^{-1}$) before and after afforestation, for each cell.

The user also specifies the number of cells to be selected or the required amount of *SYR*.

We now present more details of the two main processes in CAMF: (1) the computation of the base flow, i.e., the initial sediment accumulation matrix, denoted by *SA*, see sections 2.1-2.3, and (2) the iterative process for the identification of the best cells to perform a certain action (e.g., afforestation) in order to minimize the flow reaching the outlet, see section 2.4.

## 2.1. Sediment Accumulation

The matrices storing the local sediment production before and after afforestation are denoted respectively by $\alpha^1$ and $\alpha^2$, while $\alpha_i^k$, $k = 1, 2$ denote the local sediment production in cell $i$. The Sediment Accumulation matrix $SA$ gives for each cell $i$ the total amount of sediment $SA_i$, i.e., $\alpha_i^k$ plus the sediment received by that cell from its up-slope neighbours.

The amount of sediment that leaves cell $i$, denoted by $D_i$, is a function of $SA_i$. Currently CAMF uses a piece-wise linear convex function (Vanegas et al., 2009), depending on three parameters: retention capacity $\rho_i^k$, saturation threshold $\sigma_i^k$ and flow factor $\gamma_i^k$, where $k = 1$ if the cell is not yet afforested and $k = 2$ if the cell is afforested.

If $SA_i$ is below its retention capacity $\rho_i^k$, no sediment leaves the cell. If $SA_i \in [\rho_i^k, \sigma_i^k]$, a fraction $\gamma_i^k$ of $SA_i$ leaves the cell. If $SA_i$ is above the saturation threshold $\sigma_i^k$, the exceeding amount of sediment is fully delivered from the cell. Hence $D_i$ is modeled by

$$D_i = \begin{cases} 0, & \text{if } SA_i \leq \rho_i^k \\ \gamma_i^k(SA_i - \rho_i^k), & \text{if } \rho_i^k < SA_i \leq \sigma_i^k \\ \gamma_i^k(\sigma_i^k - \rho_i^k) + (SA_i - \sigma_i^k), & \text{if } SA_i > \sigma_i^k \end{cases} \quad (1)$$

The amount of sediment transported between cells $j$ and $i$, denoted by $D_{j,i}$, is calculated as

$$D_{j,i} = D_j \times F_{j,i} \quad (2)$$

with $D_j$ the amount of sediment that leaves cell $j$, calculated using Eq. 1 and $F_{j,i}$ the fraction of sediment flowing from cell $j$ to cell $i$.

In case of afforestation, the parameters determining the piece-wise linear convex function change when the land use of a cell is converted from its initial state (not afforested) to its new state (afforested), see Fig. 1. Afforestation decreases local sediment production and flow factor, i.e., $\alpha_i^2 < \alpha_i^1$ and $\gamma_i^2 < \gamma_i^1$, whereas it increases retention capacity and saturation threshold, i.e., $\rho_i^2 > \rho_i^1$ and $\sigma_i^2 > \sigma_i^1$. Therefore, CAMF requires two input data-sets for locally produced sediment, flow factor, retention capacity and saturation threshold, representing the catchment respectively under the original land cover (initial situation) and with every candidate cell assumed as afforested. Hence, the selection of a cell conditions the amount of sediment delivered by this cell and also conditions the amount of sediment flowing into the cells lying on the flow path(s) between this cell and the outlet(s).

Selecting cells in a raster database for maximal impact intervention

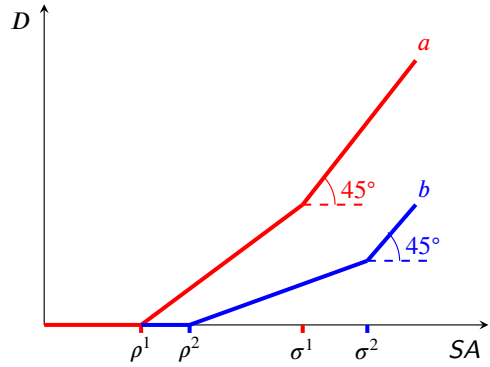

**Figure 1:** Piece-wise linear convex functions to compute the amount of sediment that leaves a cell. *SA*: sediment accumulation; *D*: amount of sediment leaving the cell. Case *a* (red curve): the cell is not afforested and *b* (blue curve): the cell is afforested; $\rho^1, \rho^2$: retention capacity before and after afforestation; $\sigma^1, \sigma^2$: saturation threshold before and after afforestation.

## 2.2. CAMF-SFD

In the original version of CAMF (Vanegas, 2010; Vanegas et al., 2012, 2014), sediment flow simulation is based on a SFD model, i.e., flow leaving a cell is assumed to be delivered entirely to the neighbouring cell(s) at lowest altitude. In particular, the Eight-Direction (D8) algorithm (O'Callaghan and Mark, 1984) is used: a cell $c$ flows to cell $i$ (one of its eight neighbors), with $i$ determined by

$$arg|min_{i=1,2,...8} Z_i \; if \; Z_c > Z_i, \tag{3}$$

where $Z_c$ is the elevation for cell $c$, see Fig. 2. If $Z_i \geq Z_c$ for all $i$, cell $c$ does not flow to a neighbor.

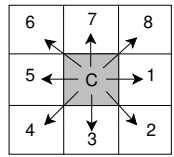

**Figure 2:** Cell $c$ and its eight neighbors.

The flow direction in CAMF-SFD can be represented as a tree that expresses the flow connectivity of the cells (Fig. 3), where each node corresponds to a raster cell, see Fig. 3b, with the root node corresponding to the outlet of the catchment and the child-parent relationship representing the sediment flow direction according to the steepest descent pathway. To compute the sediment accumulation matrix *SA*, the cells must be processed in the right order, i.e., the tree must be traversed depth-first.

As mentioned above and explained in detail in section 2.4, in the optimization iteration a single candidate cell $j$ will be (tentatively) afforested to compute the corresponding *SYR*. In this case, the *SA* must not be recomputed completely, but only the $SA_i$ values for cells $i$ on the path between cells $j$ and the root of the tree.

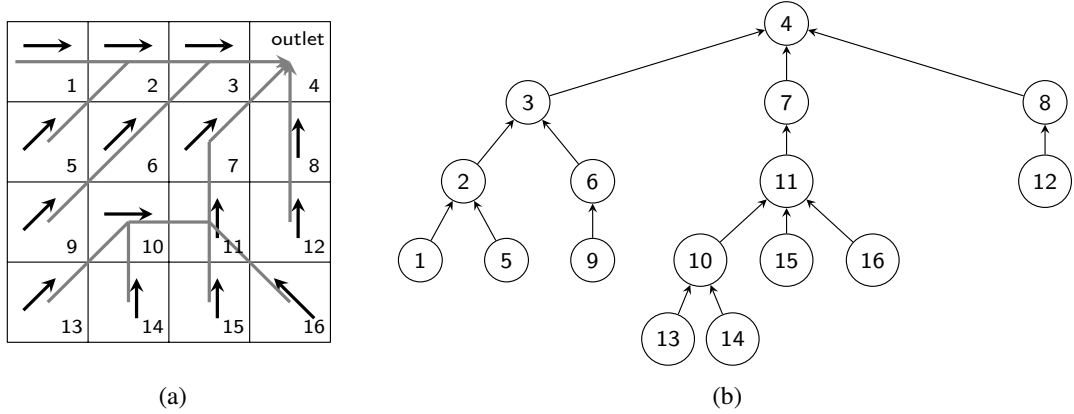

(a)    (b)

**Figure 3:** (a): Fragment of a flow direction matrix, where cell 4 represents the outlet, black arrows represent flow direction and gray arrows denote the branches of the corresponding tree representation; (b): Corresponding tree representation.

### 2.3. CAMF-MFD

However, the SFD method fails to reflect the reality of surface transport processes (Quinn et al., 1991; Wilson and Gallant, 2000). In the MFD variant of CAMF we use the Fractional Deterministic Eight-Neighbor (FD8) model (Quinn et al., 1991), but this can easily be replaced by another MFD model, e.g. the $D\infty$ algorithm (Tarboton, 1997) and the Digital Elevation MOdel Networks (DEMON) (Costa-Cabral and Burges, 1994).

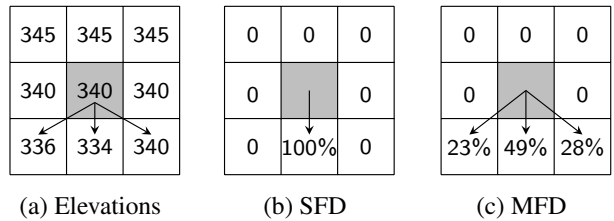

(a) Elevations    (b) SFD    (c) MFD

**Figure 4:** Difference in simulation of flow in CAMF-SFD and CAMF-MFD

In the FD8 model (Quinn et al., 1991), the proportion of sediment delivered to each down-slope neighboring cell is given by

$$F_{j,i} = \frac{\tan G_{j,i} \times L_{j,i}}{\sum_{m=1}^{n} \tan G_{j,m} \times L_{j,m}} \qquad (4)$$

with $F_{j,i}$ the fraction of sediment delivered from cell $j$ to its neighbor cell $i$; $n$ the number of down-slope neighboring cells; $G_{j,i}$ the gradient between cell $j$ and neighbor cell $i$; $L_{j,i}$ contour length: $0.5 \times$ cell size for a cardinal neighbour and $0.354 \times$ cell size for a diagonal neighbour.

In CAMF-MFD the spatial interaction and the cell connectivity is represented as a graph, instead as a tree. To construct this graph, Algorithm 1 computes for all the cells an adjacency list with ancestors and successors, based on the flow direction matrix $F$ (Figs. 5a, 5b). The ancestors of cell $i$ drain into cell $i$ and the successors of $i$ receive flow from $i$. Several methods to compute the *SA* matrix using a graph have been proposed, see e.g. (Qin and Zhan, 2012; Jiang et al., 2013). However, processing the graph is very memory and compute intensive for large data-sets. In the minimization process in CAMF, we compute the *SA* matrix many times. Therefore, in a pre-processing step Algorithm 2 sorts the cells by a topological sorting algorithm (Kahn, 1962), as in Anand et al. (2020), so that for each ancestor–successor pair, the ancestor always appears before its successor in the sorted list (Fig. 5c). The computation of *SA* is then performed cell per cell in the order given by the sorted list, from left to right, see Algorithm 3.

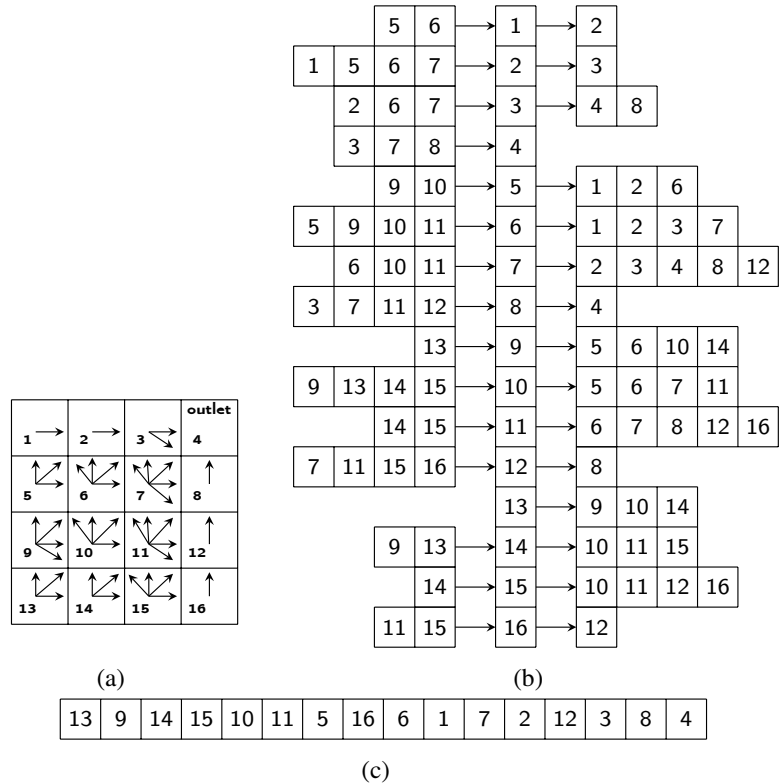

**Figure 5:** Representation of flow direction in CAMF-MFD. (a) Input raster flow direction matrix $F$; (b) Adjacency list representing the spatial interaction graph; (c) Topologically sorted list of cells.

As in the case of SFD, *SA* must not be completely recomputed when a single candidate cell $j$ is (tentatively) afforested to compute the corresponding *SYR*. However, in case of MFD, $SA_i$ for all cells $i$ on all paths between cell $j$ and the outlet must be recomputed, which corresponds to a substantially larger part of the *SA* matrix compared to the single path used in SFD. Also, recomputing $SA_i$ for only these cells $i$ requires to store and process the complete graph. Experiments indicated that this is very expensive. Therefore, when cell $j$ is (tentatively) afforested, $SA_i$ is recomputed for all cells $i$ in the sorted list between cell $j$ and the outlet.

### 2.4. Minimizing sediment yield at the outlet

Each iteration of the CAMF algorithm consists of the following three steps:

1. Each candidate cell not yet selected in previous iterations is tentatively selected and the *SA* matrix is recomputed

---

**Algorithm 1** Create an adjacency list from flow direction matrix

**Input:** MFD matrix $F$

1. Create data structure to store adjacency list $A$

**for** each cell $i$ of $F$ **do**
    **for** each neighbor $j$ of $i$ **do**
        **if** $i$ drains into $j$ **then**
            2. Add $j$ as successor of $i$ in $A$
            3. Add $i$ as ancestor of $j$ in $A$
        **end if**
    **end for**
**end for**

**Output:** Adjacency list $A$

---

**Algorithm 2** Sort the adjacency list

**Input:** Adjacency list $A$

1. Create data structure $S$ to contain the sorted list
2. Create data structure $E$ to store cells without incoming ancestors

**for** each cell $i$ in $A$ without ancestors **do**
    3. Store $i$ in $E$
**end for**

**while** $E$ is non-empty **do**
    4. Get cell $i$ from $E$
    5. Remove cell $i$ from $E$
    6. Add $i$ to $S$
    **for** each cell $j$ in $A$ with $i$ as ancestor **do**
        7. Remove $i$ from ancestors of $j$ in $A$
        **if** $j$ has no other ancestors **then**
            8. Add $j$ to $E$
        **end if**
    **end for**
**end while**

**Output:** Sorted list $S$

---

to evaluate the effect of selecting this cell. In section 2.1 we indicated that when a cell is selected for afforestation the amount of sediment delivered to its neighbours decreases. This decrease is propagated to the outlet and results in a decrease in *SY* and thus an increase in the *SYR*.

Let $SY^b$ be the *SY* in the base flow; $SY_i^k$ the *SY* if cell $i$ would be afforested in iteration $k$; $SYR_i^k$ the total *SYR* if cell $i$ would be afforested in iteration $k$. Then

$$SYR_i^k = SY^b - SY_i^k \tag{5}$$

---

---

**Algorithm 3** Compute Sediment Accumulation *SA*

**Input:** Sorted list $S$, adjacency list $A$, local sediment production $\alpha^k$, flow factor $\gamma^k$, retention capacity $\rho^k$, saturation threshold $\sigma^k$, with $k$ indicating the values before afforestation ($k = 1$) or after afforestation ($k = 2$)

1. Create data structure *SA* to store the sediment accumulation for every cell

**for** each cell $i$ in $S$ **do**
   2. $SA_i \leftarrow \alpha_i^k$
   **for** each ancestor $j$ of $i$ in $A$ **do**
     **if** $SA_j < \rho_j^k$ **then**
       3. $D_j \leftarrow 0$
     **else**
       **if** $SA_j \geq \rho_j^k$ and $SA_j \leq \sigma_j^k$ **then**
         4. $D_j \leftarrow \gamma_j^k (SA_j - \rho_j^k)$
       **else**
         **if** $SA_j > \rho_j^k$ **then**
           5. $D_j \leftarrow \gamma_j^k (\sigma_j^k - \rho_j^k) + (SA_j - \sigma_j^k)$
         **end if**
       **end if**
     **end if**
     6. $D_{j,i} \leftarrow D_j \times F_{j,i}$
     7. $SA_j \leftarrow SA_j - D_{j,i}$
     8. $SA_i \leftarrow SA_i + D_{j,i}$
   **end for**
**end for**

**Output:** Sediment accumulation *SA*

---

2. Cells are ranked in descending order according to their $SYR_i^k$ value.

3. The cell(s) with the maximal *SYR* value are added to the set of cells selected for afforestation.

The iterative procedure ends when the number of selected cells reaches a user-specified value (see Algorithm 4). Note that this stop criterion can easily be replaced by a test on achieving a user-specified *SYR*.

Algorithm 3 for computing *SA* has a linear temporal complexity in the number of cells. In the worst case, Algorithm 4 has a quadratic temporal complexity in function of the number of cells, since the number of candidate cells can be of the order of the number of cells. In addition, most probably only one cell is selected per iteration. Hence, the total temporal complexity of CAMF-MFD is proportional to the total number of cells in the data-set, the number of candidate cells and the number of cells selected for afforestation.

## 3. Case studies

### 3.1. Tabacay catchment

The first case study deals with the Tabacay river basin in Ecuador. It is a relevant region for afforestation since the Tabacay river provides drinking water to the city of Azogues. The area of the Tabacay catchment is 6 639 *ha*,

---

**Algorithm 4** Determine the cells to be selected

**Input:** Number of cells to be selected $n$

1. Create data structure $S$ to store cells to be selected
2. $S \leftarrow \emptyset$
3. $k \leftarrow 0$

**while** size of $S < n$ **do**
    4. $k \leftarrow k + 1$
    **for** each candidate cell $i$ not yet selected **do**
        5. Compute $SA$ matrix and $SYR_i^k$ by simulating the afforestation of cell $i$
    **end for**
    6. Rank cells according to $SYR$
    7. Put cell(s) with highest $SYR$ in solution set $S$
**end while**

**Output:** Set of selected cells $S$

---

with altitude from 2 482 $m$ to 3 731 $m$ above sea level. Agriculture and pasture cover 2 400 $ha$ (39% of the total area) (Wijffels and Van Orshoven, 2009). The widespread agricultural land use, even on steep slopes, leads to large amounts of sediment produced and transported towards the river, which results in severe land, river and reservoir degradation. A raster geo-database representing the Tabacay catchment with a resolution of $30m \times 30m$ is used, containing 122 830 cells, of which 73 471 are active cells (with actual values). Only part of the cells under agriculture and pasture are considered as candidate cells, i.e., cells that can be selected for afforestation (27 246 cells).

The initial sediment production map $\alpha^1$ was computed using the Revised Universal Soil Loss Equation (RUSLE), Eq. 6 proposed by Renard et al. (1991).

$$E = R \times K \times LS \times C \times P \tag{6}$$

where $E$: annual soil loss ($ton\, ha^{-1}\, yr^{-1}$); $R$: rainfall erosivity factor ($MJ\, mm\, ha^{-1}\, h^{-1}\, yr^{-1}$); $K$: soil erodibility factor ($ton\, h\, MJ^{-1}\, mm^{-1}$); $LS$: slope length and slope steepness factors; $C$: cover management factor; $P$: support practice factor.

The values $P = 1$ and $R = 1\,599\, MJ\, mm\, ha^{-1}\, h^{-1}\, yr^{-1}$ were obtained from Cisneros Espinosa et al. (1999). The equation introduced by Wischmeier (1978) was used to calculate the $K$-factor from soil granulometric fractions, see Table 1. The $C$-factor map was generated by assigning the C-values presented in Table 2, taken from Estrella (2015), to the land cover map (Fig. 7a). The $LS$-factor map was computed from the DEM using a SAGA-GIS tool (Conrad, 2003), based on the slope and specific catchment area, following the approach of Desmet and Govers (1996). The $LS$-factor map and the initial sediment production $\alpha^1$ are shown in Figs. 7b, 7c.

The initial flow factor map $\gamma^1$ in Eq. 1 was computed by a linear transformation of the original topographic slope map using min-max normalization (Han et al., 2012). The values of the other parameters of Eq. 1, i.e., $\rho^1$, $\rho^2$, $\sigma^1$, $\sigma^2$, $\alpha^2$ and $\gamma^2$, listed in Table 3, are taken from Estrella (2015). Note that Estrella (2015) used a slightly different initial local sediment production map $\alpha^1$, since the $LS$-factor was computed differently.

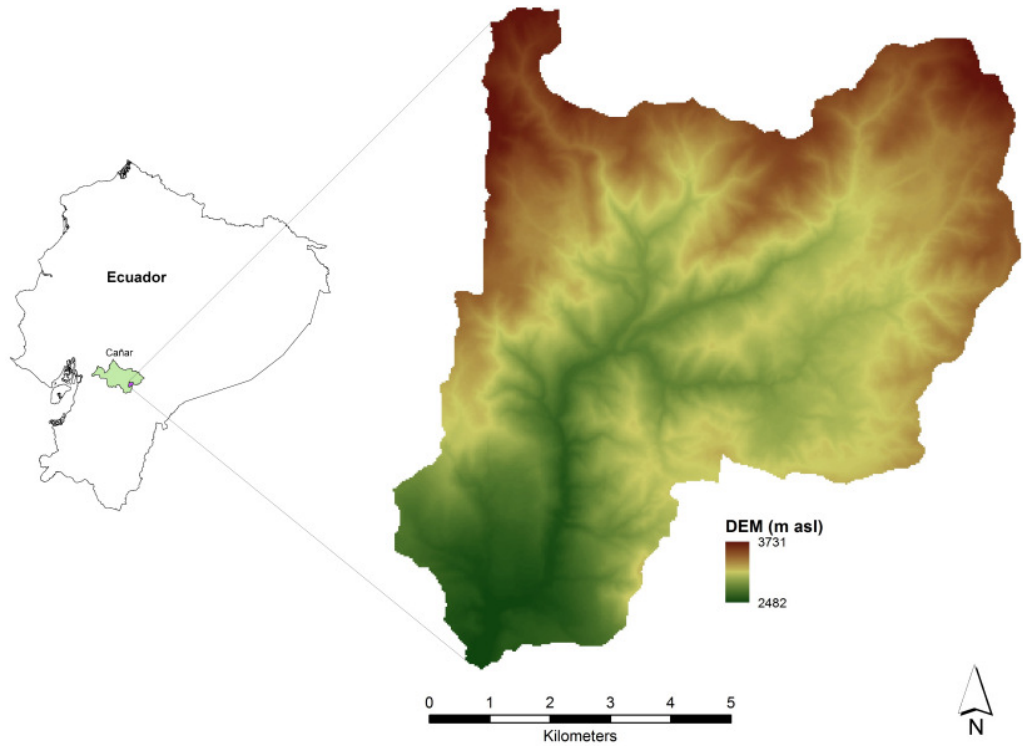

**Figure 6:** Digital Elevation Model of the Tabacay catchment and its location in Ecuador (Estrella et al., 2014a).

**Table 1**
K-factor calculated for each soil type in Tabacay catchment (Estrella, 2015).

| Soil Type | Soil Texture | K-factor |
|---|---|---|
| Umbric Leptosol | Sandy clay loam | 0.0397 |
| Umbric Andosol | Loam | 0.0373 |
| Dystric Cambisol | Loam | 0.0376 |
| Ferralic Cambisol | Clay loam | 0.0253 |
| Calcaric Regosol | Sandy loam | 0.0452 |
| Calcaric Cambisol | Sandy clay loam | 0.0290 |
| Eutric Cambisol | Silt loam | 0.0518 |
| Eutric Regosol | Sandy loam | 0.0521 |

Additionally, two smaller regions around the outlet of the catchment were cropped from the original data-set, to evaluate the scalability of both CAMF variants. Figs. 8a, 8b show the corresponding fragments of the initial local sediment production map $\alpha^1$.

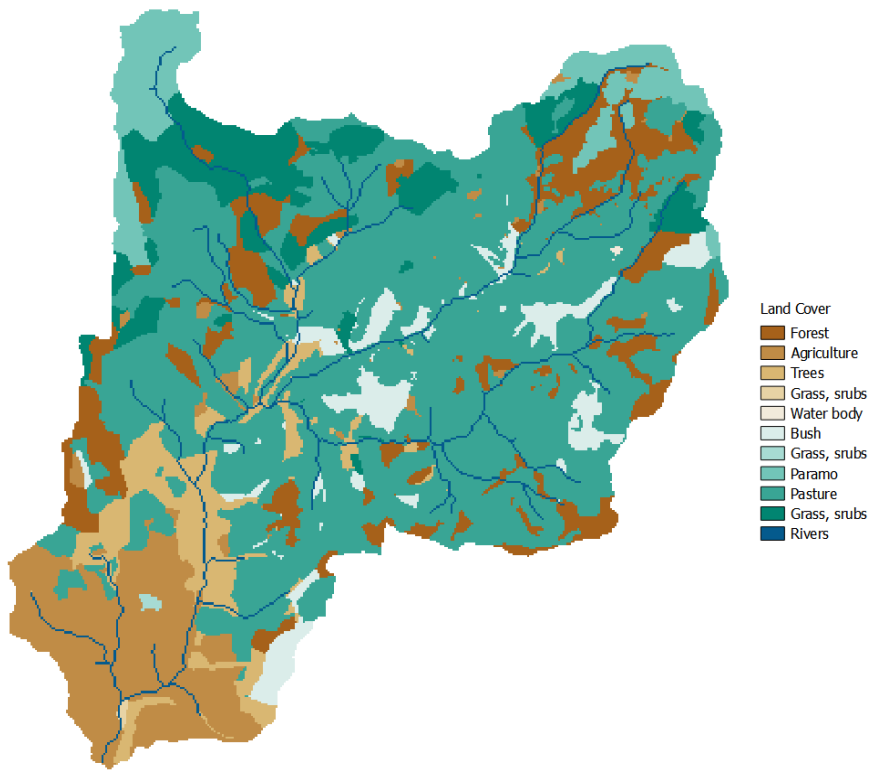

(a)

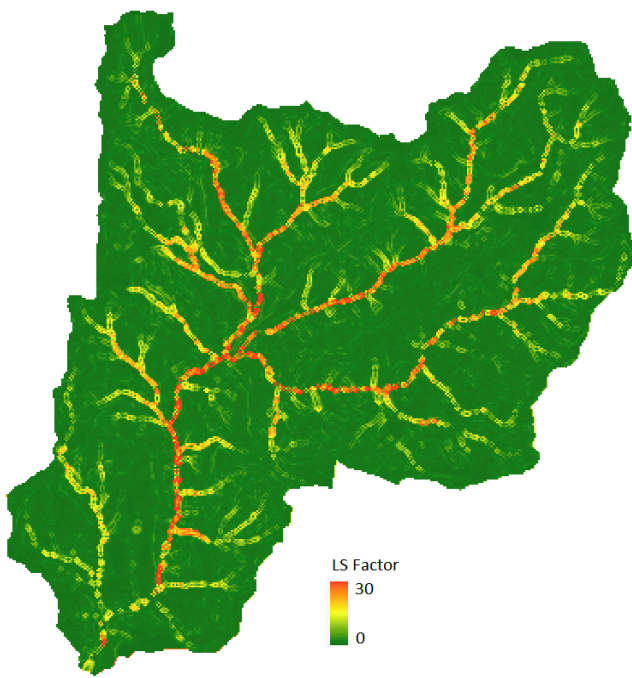

(b)

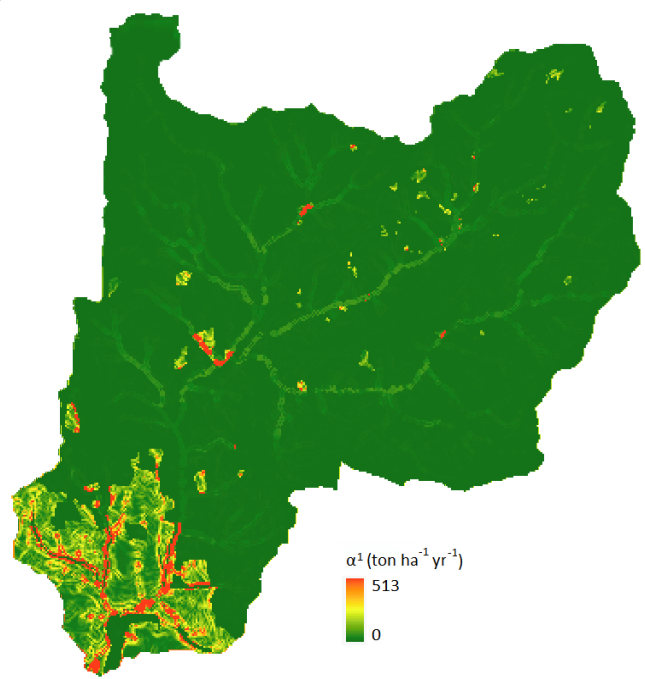

(c)

**Figure 7:** (a): Land cover map of the Tabacay catchment; (b): $LS$-factor map used to compute the initial local sediment production map $\alpha^1$; (c): Initial local sediment production map $\alpha^1$ ($ton\,ha^{-1}\,yr^{-1}$) calculated by means of RUSLE using Eq. 6.

Selecting cells in a raster database for maximal impact intervention

**Table 2**
C-factor values assigned to each land use type in Tabacay catchment (Estrella, 2015).

| Land cover type | C-factor |
|---|---|
| Agriculture | 0.2 |
| Forest | 0.001 |
| Lake | 0 |
| Bush | 0.003 |
| Pasture | 0.003 |
| Wetlands non-afforested | 0.003 |
| Other features | 0.003 |

**Table 3**
Parameter values used in the experiments with CAMF-SFD and CAMF-MFD for the Tabacay data-set.

| Parameter | Initial value (before afforestation) | Second value (after afforestation) |
|---|---|---|
| Sediment production | Calculated by RUSLE, Fig. 7c | $\alpha^2 = 0.83 \times \alpha^1$ |
| Retention capacity | $\rho^1 = 0.37 \times \alpha^1$ | $\rho^2 = 0.61 \times \alpha^1$ |
| Saturation threshold | $\sigma^1 = 0.96 \times \alpha^1$ | $\sigma^2 = 0.98 \times \alpha^1$ |
| Flow Factor | Normalized slope from DEM | $\gamma^2 = 0.75 \times \gamma^1$ |

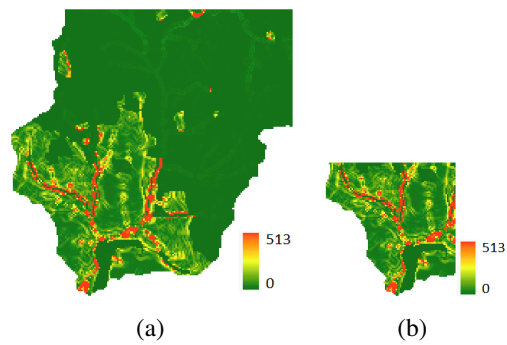

(a)      (b)

**Figure 8:** Maps cropped from the initial local sediment production map $\alpha^1$ ($ton\,ha^{-1}\,yr^{-1}$) in the Tabacay catchment. (a): $\frac{1}{4}$ of the original; (b): $\frac{1}{16}$ of the original.

### 3.2. Maarkebeek catchment

The second case study deals with the Maarkebeek river basin (Belgium) with an area of $\approx 4\,800\,ha$, with altitudes from $14.1m$ to $146.9m$, i.e. a rather flat area, containing mostly agricultural area, dominated by arable land, with $\approx 10\%$ urbanized and $\approx 10\%$ afforested Gabriels et al. (2022). The Maarkebeek geo-database with a resolution of $20m \times 20m$ contains 129 097 active cells, of which 53 792 are candidate cells for afforestation (cells under agriculture and pasture), see the DEM in Figure 9. Since the Maarkebeek data-set is larger than the Tabacay data-set the computational cost is substantially higher.

As in Tabacay catchment, the initial local sediment production map $\alpha^1$ (Fig. 11) was generated by means of RUSLE, Eq. 6. The values $P = 1$, $R = 870\,MJ\,mm\,ha^{-1}\,h^{-1}\,yr^{-1}$, the $K$-factor associated to each soil type (Table 4) and the $C$-factor assigned to each land use type (Table 5), were obtained from Deproost et al. (2018). The $LS$-factor map was computed using the same tool as for the Tabacay catchment. The values for $\rho^1$, $\rho^2$, $\sigma^1$, $\sigma^2$, $\alpha^2$ and $\gamma^2$ in Table 6 were calibrated such that *SY* in the base flow for the SFD model is close to the value given in Deproost et al. (2018).

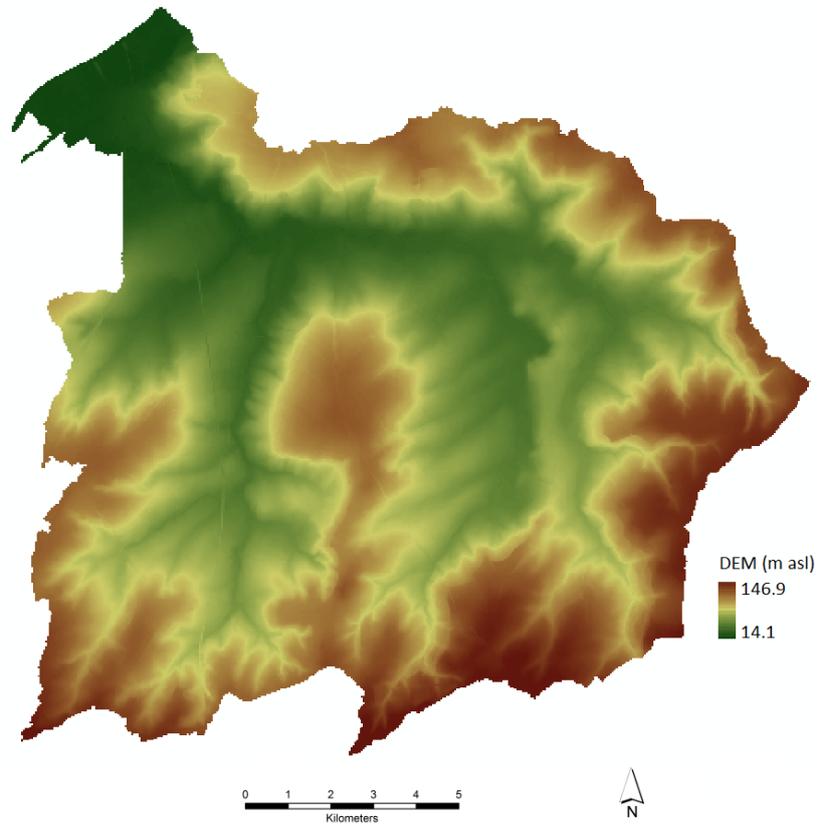

**Figure 9:** Digital Elevation Model of the Maarkebeek catchment in Belgium (Gabriels et al., 2022).

**Table 4**
K-factor calculated for each soil type in Maarkebeek catchment.

| Soil Texture | K-factor |
|---:|---:|
| Silty sand | 0.02 |
| Light sandy loam | 0.025 |
| Sand loam | 0.4 |
| Clay | 0.042 |

**Table 5**
C-factor values assigned to each land use type in Maarkebeek catchment.

| Land cover type | C-factor |
|---:|---:|
| Lake and Infrastructure | 0 |
| Grass, shrubs (by the side of the road) | 0.001 |
| Forest | 0.01 |
| Agricultural, pasture and arable land | 0.37 |

## 4. Results

All experiments have been performed on a Xeon E5-2697 v3 CPU (2.6 GHz) with 28 cores (2 sockets, each with 14 cores) and 128 GB memory, with Ubuntu Bionic Linux (18.044.15.0-147-generic x86 64 kernel) as Operating

Selecting cells in a raster database for maximal impact intervention

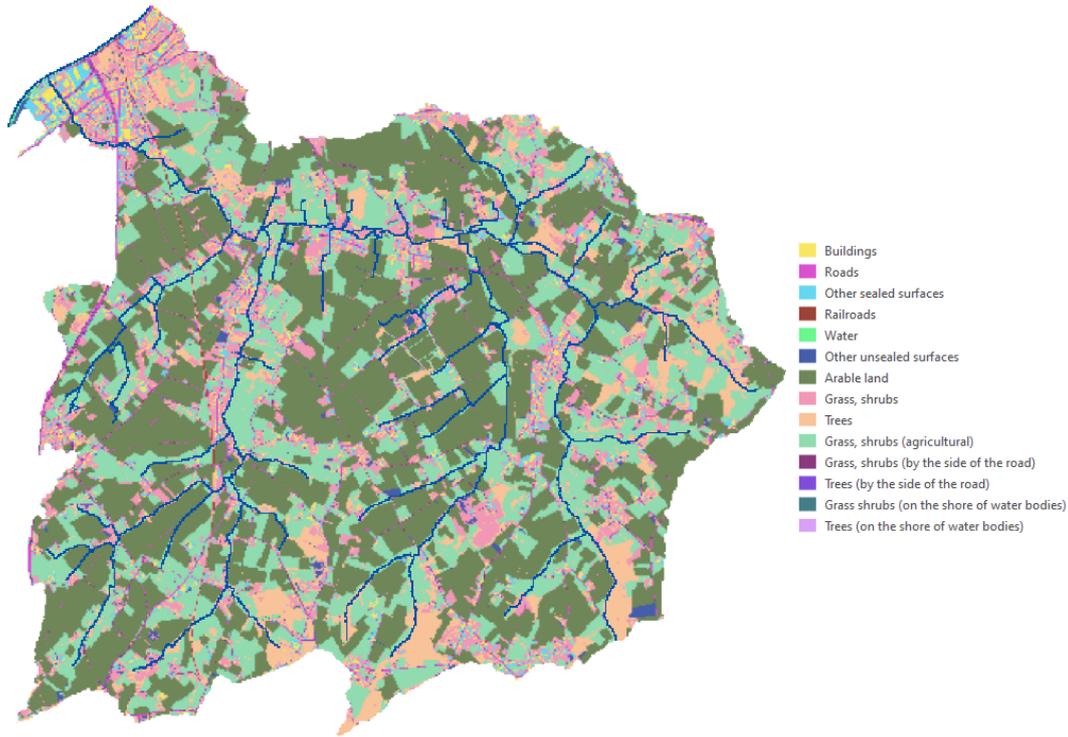

**Figure 10:** Land cover map of the Maarkebeek river catchment (Gabriels et al., 2022).

**Table 6**
Parameter values used in the experiments with CAMF-SFD and CAMF-MFD for the Maarkebeek data-set.

| Parameter | Initial value (before afforestation) | Second value (after afforestation) |
| --- | --- | --- |
| Sediment production | Calculated by RUSLE, Fig. 11 | $\alpha^2 = 0.83 \times \alpha^1$ |
| Retention capacity | $\rho^1 = 0.55 \times \alpha^1$ | $\rho^2 = 0.73 \times \alpha^1$ |
| Saturation threshold | $\sigma^1 = 1 \times \alpha^1$ | $\sigma^2 = 1.02 \times \alpha^1$ |
| Flow Factor | Normalized slope from DEM | $\gamma^2 = 0.75 \times \gamma^1$ |

System. Although we have performed the experiments on Linux, the implementation of CAMF in C++ can be used under Linux/Unix and Windows.

### 4.1. Sediment yield reduction with CAMF-SFD and CAMF-MFD

Fig. 12 shows for the Tabacay catchment the *SA* matrices for the base flow, computed with CAMF-SFD and CAMF-MFD. With MFD, the maximum $SA_i$ is lower.

Table 7 shows that, for a given percentage of afforested cells, CAMF-MFD predicts lower *SYR* values than CAMF-SFD, but the ratio *SYR* over $SY^b$, denoted by "% SYR", is higher. Indeed, MFD increases the SI among cells and more cells are involved in the distribution of sediment. Therefore more cells retain (part of the) sediment and less sediment flows to the outlet. Further, in both cases, the cells selected first (up to 5%) contribute most to the *SYR* and the extra gain decreases with increasing number of afforested cells. By afforesting 20% of the candidate cells the maximum attainable *SYR* is nearly reached, as indicated by the horizontal lines in Fig. 13a. Due to the values that we assigned to $\rho^1$, $\rho^2$, $\sigma^1$, $\sigma^2$, $\lambda^1$ and $\lambda^2$, we observed that after afforesting the required number of cells, the *SYR* value increases, which indeed is in agreement with the fact that afforestation is an effective means in this regard (Costin, 1980; Heil et al., 2007; Nearing et al., 2005).

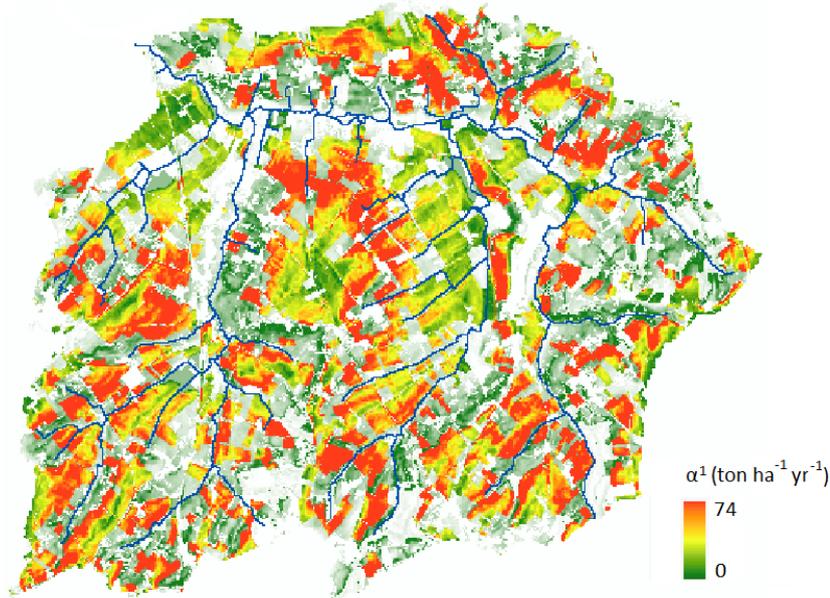

**Figure 11:** Initial local sediment production map $\alpha^1$ ($ton\,ha^{-1}\,yr^{-1}$) of Maarkebeek catchment calculated by means of RUSLE using Eq. 6.

Fig. 13a shows the *SYR* in function of the number of cells selected for afforestation. The extra gain in *SYR* per 1000 selected cells decreases, and becomes very small after afforesting 3000 cells, confirming that CAMF first selects the cells leading to the highest *SYR*. We observe that the MFD model predicts a larger relative *SYR* by afforesting the first 1000 cells than the SFD model by afforesting 8000 cells, which is a significant observation, considering that we expect that it reflects more closely the reality.

**Table 7**
*SY* and *SYR* obtained with CAMF-SFD and CAMF-MFD for the Tabacay data-set. Afforestation of up to 30% of the candidate cells.

| *SY* Base flow ($ton\,yr^{-1}$) | | # Afforested cells | *SYR* ($ton\,yr^{-1}$) | | % *SYR* | |
|---|---|---|---|---|---|---|
| SFD | MFD | | SFD | MFD | SFD | MFD |
| 22 209 | 13 691 | 5% = 1 362 | 5 326.60 | 4 227.82 | 23.98 | 30.88 |
| | | 10% = 2 724 | 5 803.09 | 4 855.23 | 26.13 | 35.46 |
| | | 20% = 5 448 | 6 034.20 | 5 256.75 | 27.17 | 38.40 |
| | | 30% = 8 172 | 6 136.12 | 5 358.93 | 27.63 | 39.14 |

Fig. 14 shows the selected cells by CAMF-MFD for afforestation of up to 30% of the candidate cells in the Tabacay catchment. Note that the first cells selected (5% and 10%) are concentrated in areas with high sediment production, as already observed for CAMF-SFD (Vanegas et al., 2012; Estrella et al., 2014b). For example, the 11 candidate cells *i*

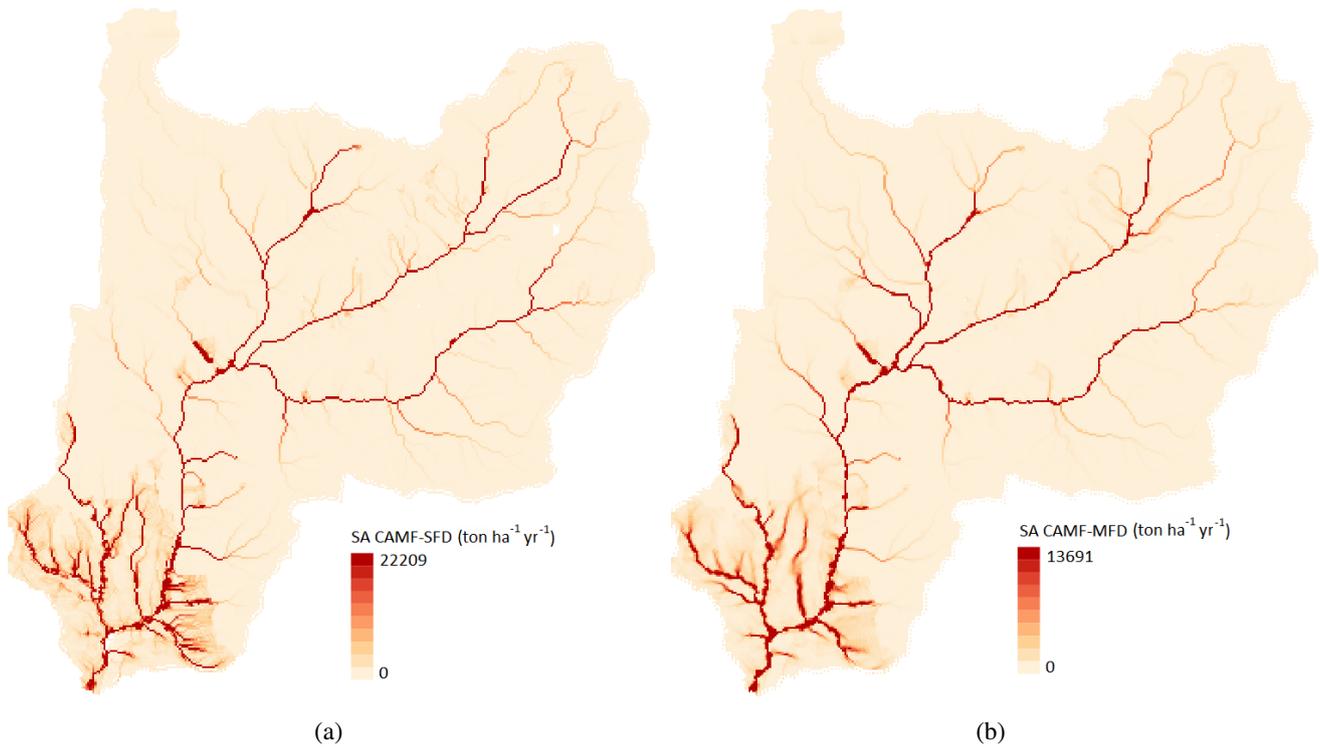

**Figure 12:** Representation of $SA$ ($ton\,ha^{-1}\,yr^{-1}$) for the base flow in Tabacay catchment. (a): CAMF-SFD and (b): CAMF-MFD.

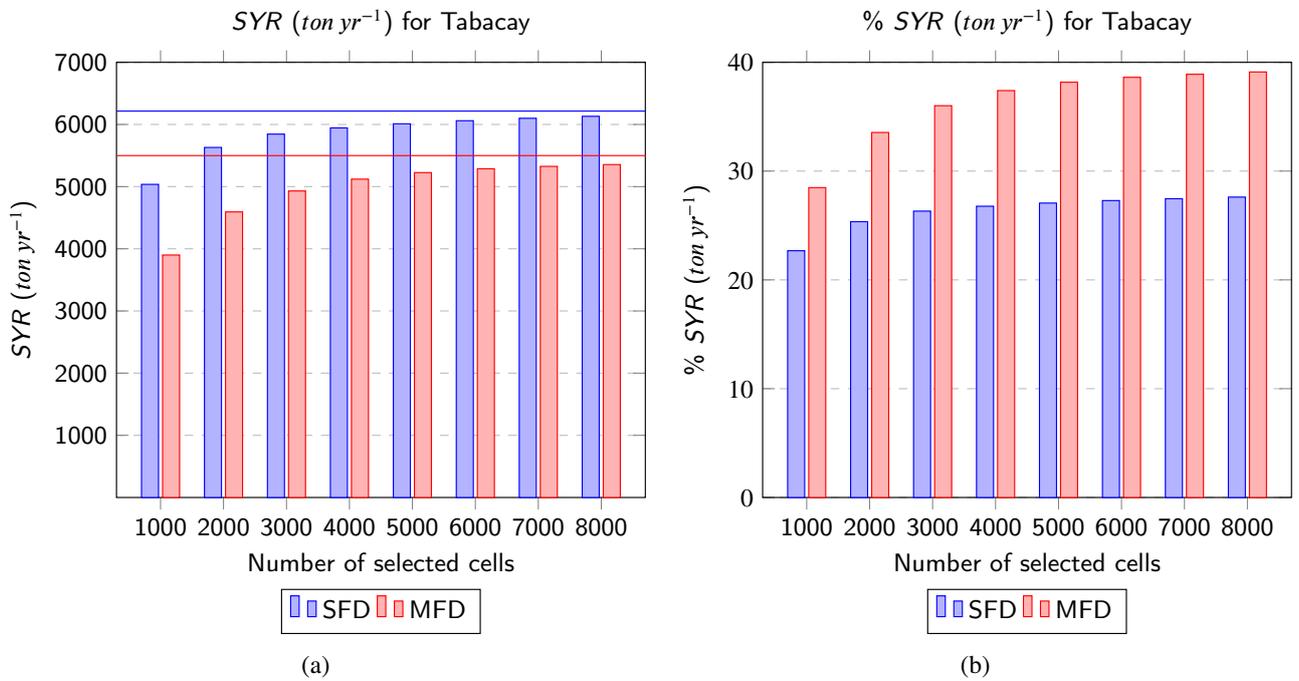

**Figure 13:** Evolution of the $SYR$ ($ton\,yr^{-1}$) for Tabacay data-set using CAMF-SFD and CAMF-MFD. (a): Absolute reduction. Horizontal lines: maximum attainable $SYR$ by afforesting all candidate cells; (b): Relative reduction.

with $\alpha_i^2 > 250\,ton\,ha^{-1}\,yr^{-1}$ were selected in the first 25 iterations.

The results for the Maarkebeek catchment are presented in Table 8 and Figs. 15a, 15b. We observe some differences with the results for Tabacay: (1) the MFD model predicts a larger *SY* in the base flow than the SFD model. (2) When only 5% of the cells is afforested, the *SYR* is close to the maximum attainable value for the SFD model, while Fig. 15a shows that this is not the case for the MFD model. (3) If less than 2% of the candidate cells is selected, the SFD model predicts a larger relative *SYR* value than MFD. On the other hand, also in this case the first cells selected are concentrated in areas with high sediment production.

**Table 8**
*SY* and *SYR* obtained with CAMF-SFD and CAMF-MFD for the Maarkebeek data-set. Afforestation of up to 5% of the candidate cells.

| *SY* Base flow ($ton\,yr^{-1}$) | | # Afforested cells | *SYR* ($ton\,yr^{-1}$) | | % *SYR* | |
|---|---|---|---|---|---|---|
| SFD | MFD | | SFD | MFD | SFD | MFD |
| | | 1% = 538 | 1 753.77 | 2 087.95 | 23.72 | 21.42 |
| | | 2% = 1 076 | 2 479.67 | 3 277.11 | 33.54 | 33.62 |
| 7 392 | 9 747 | 3% = 1 614 | 2 697.88 | 4 038.57 | 36.50 | 41.43 |
| | | 4% = 2 152 | 2 750.32 | 4 534.70 | 37.21 | 46.52 |
| | | 5% = 2 687 | 2 773.61 | 4 825.48 | 37.52 | 49.50 |

## 4.2. Scalability and computational cost of CAMF-MFD

We study the scalability of the CAMP-SFD and CAMF-MFD algorithms by measuring the execution times for the Tabacay data-sets. The sequential execution times are shown in Table 9.

The base flow is computed by calculating the *SA* matrix. Hence, the number of operations is proportional to the number of active cells. While the number of cells increases by a factor of 4 and 13 from the small to the intermediate and the original data-set, the execution times grow much slower. On current computers, execution times are mainly determined by the cost of data access, from main memory to caches and processor. Note that calculating *SA* uses a very irregular data access pattern. Probably caching works more effectively for larger data-sets. Of course, the execution time of CAMF-MFD is higher than of CAMF-SFD due to the increased SI among cells. To tentatively afforest each (not yet selected) candidate cell $j$ in each iteration of CAMF, the *SA* matrix is recomputed. As mentioned in section 2.4, this requires to recompute the $SA_i$ values for cells $i$ appearing after cell $j$ in the sorted list $S$. Thus the number of operations in one iteration is approximately $0.5 \times n$ larger than the time to compute the base flow, with $n$ the number of candidate cells. However, the ratio of the execution times for one iteration and for the computation of the base flow is much smaller for the small and intermediate data-sets, while for CAMF-MFD with the original data-set the ratio is $\approx n$. This again shows that the number of operations is a poor indicator of the execution time on current computers.

In the optimal implementation of CAMF-SFD (Estrella, 2015), the flow connectivity of the cells is represented as a tree where the root corresponds to the outlet of the catchment and child nodes deliver sediment to their parents. Note that, in this case, only the cells on the path between cell $j$ and the outlet (one branch of the tree) must be recomputed. But this optimization is not implemented in CAMF-MFD. However, with MFD, the SI among cells increases and the data structure used to model the flow paths becomes a graph, significantly more complex than the tree in CAMF-SFD, which leads to a substantial increase in the computational cost of the algorithm.

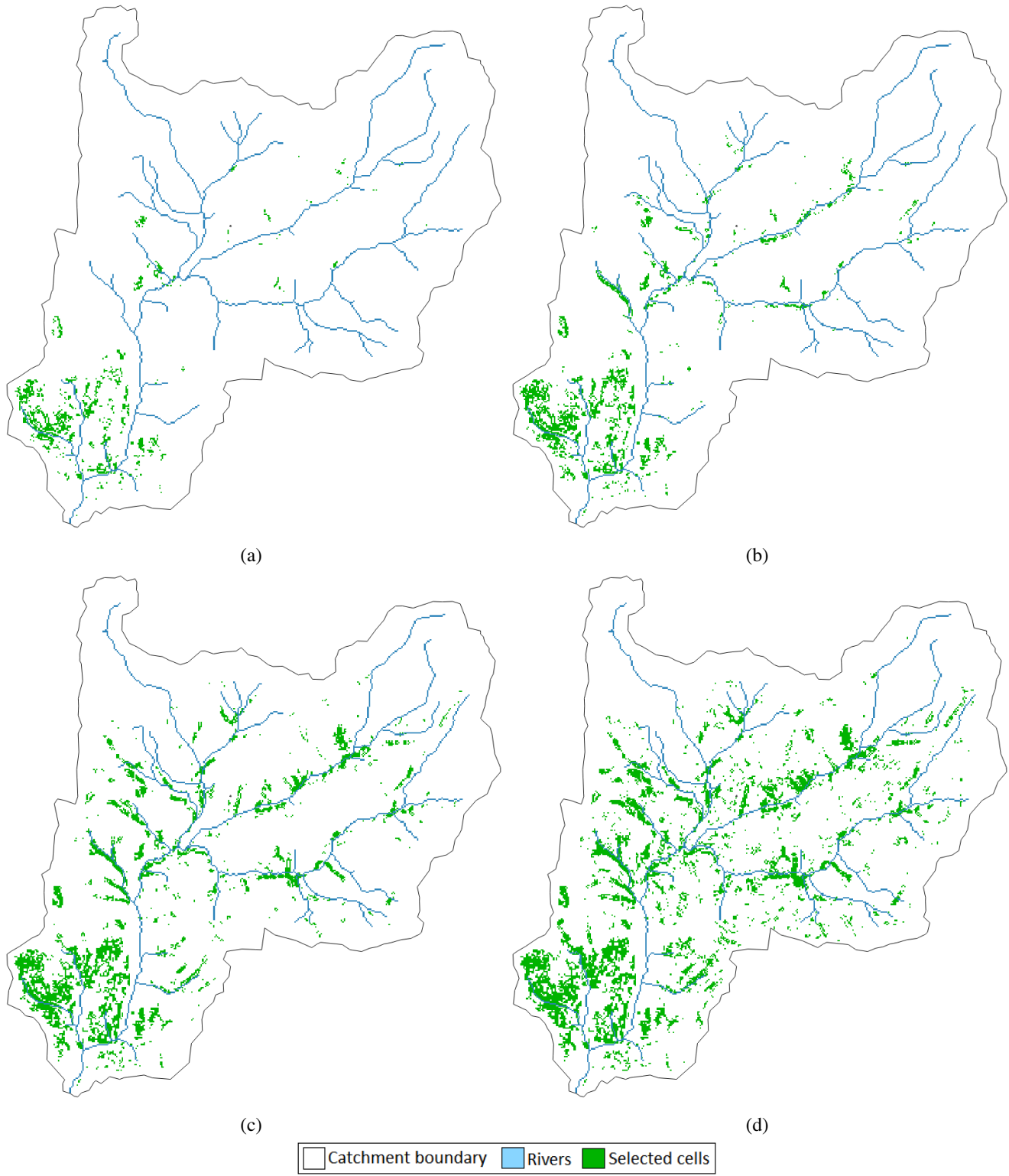

**Figure 14:** Areas selected by CAMF-MFD for afforestation in Tabacay catchment. (a): 5% of total candidate cells; (b): 10% of total candidate cells; (c): 20% of total candidate cells; (d): 30% of total candidate cells.

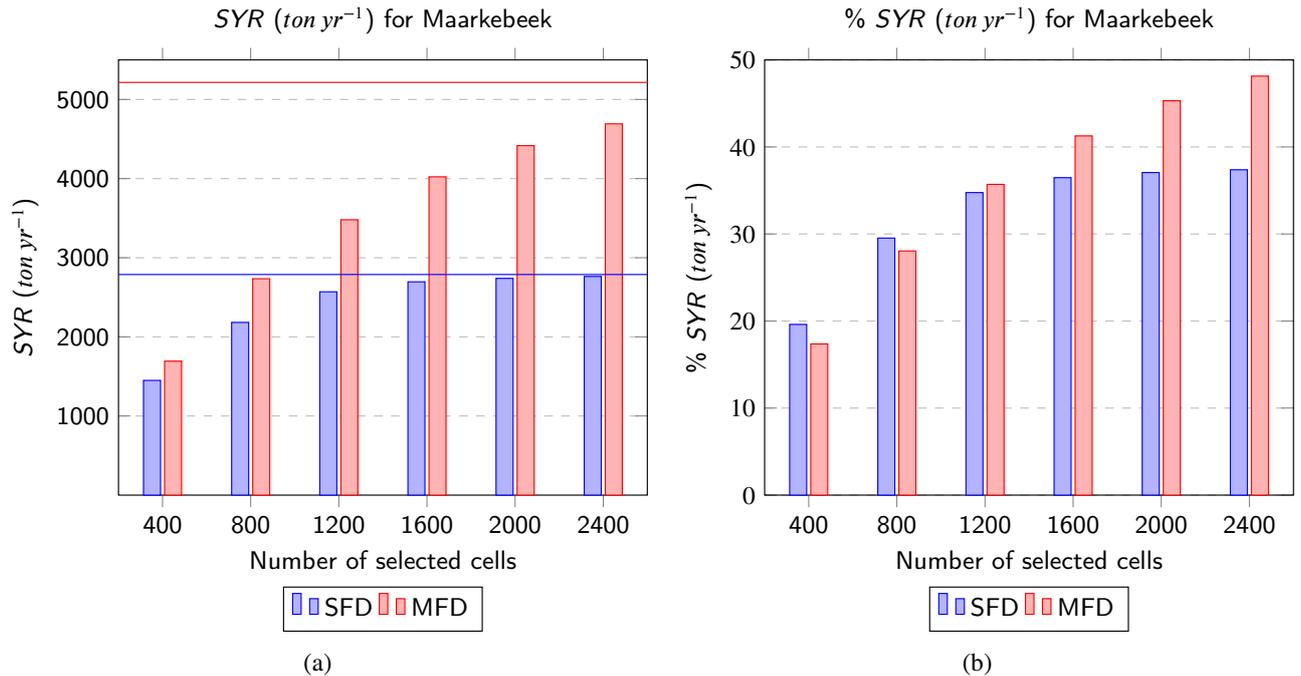

Figure 15: Evolution of the $SYR$ ($ton\ yr^{-1}$) for Maarkebeek data-set using CAMF-SFD and CAMF-MFD. (a): Absolute reduction. Horizontal lines: maximum attainable $SYR$ by afforesting all candidate cells; (b): Relative reduction.

**Table 9**
CPU times for the calculation of the $SA$ matrix in the initial situation and for one iteration, using the Tabacay data-set. Small and intermediate data-sets correspond to $\frac{1}{16}$ and $\frac{1}{4}$ of the original respectively; Ratio 1: ratio of intermediate to small data-set; Ratio 2: ratio of original to small data-set.

| Data-set | Small | | Intermediate | | Ratio 1 | | Original | | Ratio 2 | |
|---|---|---|---|---|---|---|---|---|---|---|
| Dimensions | 89 x 87 | | 178 x 173 | | | | 355 x 346 | | | |
| # cells | 7 743 | | 30 794 | | 3.98 | | 122 830 | | 15.86 | |
| # active cells | 5 475 | | 22 494 | | 4.11 | | 73 471 | | 13.42 | |
| # candidate cells | 2 259 | | 9 859 | | 4.36 | | 27 246 | | 12.06 | |
| | CPU time (s) | | CPU time (s) | | Ratio | | CPU time (s) | | Ratio | |
| | SFD | MFD | SFD | MFD | SFD | MFD | SFD | MFD | SFD | MFD |
| Base Flow SA | 0.003 | 0.018 | 0.006 | 0.023 | 2.19 | 1.28 | 0.009 | 0.028 | 3.45 | 1.56 |
| 1 iteration | 0.476 | 1.373 | 9.658 | 26.669 | 20.28 | 19.43 | 110.819 | 651.202 | 232.72 | 474.33 |

## 5. Conclusions

The raster-based CAMF method selects optimal sites for implementing an action, e.g., afforestation, to minimize sediment delivery at the outlet of a river catchment. Sediment transport among cells is modeled by a piece-wise linear function, parameterized by retention capacity, saturation threshold and a flow factor. The original CAMF uses a SFD method for flow simulation. We described and analyzed the integration in CAMF of a MFD method, in which sediment leaving a cell flows in all down-slope directions, allowing more realistic flow simulations.

The various algorithmic steps are presented in detail, highlighting the increased complexity of the sediment transport simulation in the MFD variant. The CAMF-MFD implementation in C++ is applied for afforestation planning in two river catchments: Tabacay catchment in Ecuador and Maarkebeek catchment in Flanders, Belgium.

The results show that CAMF-MFD iteratively selects, from a large set of candidate cells and in the presence of spatial interaction, those cells for which the marginal contribution of afforestation to the sediment yield reduction is highest. A major characteristic of the selected cells is that they lie in regions with a high local sediment production. CAMF-MFD predicts a lower sediment yield at the outlet of the catchment than CAMF-SFD for the Tabacay catchment with large elevation differences, while the opposite is true for the flat Maarkebeek catchment. When a rather large number of cells is afforested, MFD predicts a larger relative sediment yield reduction, due to the retention capacity of the many cells on the path between the afforested cell and the outlet.

CAMF-MFD has a substantially larger computational cost than CAMF-SFD, since updating the sediment accumulation matrix after afforesting a single cell is much more expensive in CAMF-MFD than in CAMF-SFD. Therefore the execution time of each iteration in the sediment minimization process is much larger in CAMF-MFD. Hence there is a clear need to exploit parallelization and to reduce the cost of each iteration and also the number of iterations. This topic is addressed in a forthcoming paper, where we show that execution time can substantially be reduced.

## 6. Acknowledgments

The authors would like to acknowledge the VLIR-UOS projects "Networks 2019 Phase 2 Cuba ICT" (grant for PhD research of G. Castillo Reyes) and "Global Minds KU Leuven" (Short term Research Stay of R. Estrella).

**Code availability section**

Name of the code/library: A-CAMF

Developer: Grethell Castillo Reyes

Contact: Department of Computer Science, Celestijnenlaan 200A box 2402, 3001 Leuven, Belgium; Data Representation and Analysis Center, San Antonio de los Baños Km 2½, University of Informatic Sciences, Cuba.

Email: grethell.castilloreyes@kuleuven.be, gcreyes@uci.cu. Tel.: +32496998509

Year First Available: 2022

Hardware requirements: Code was tested on a a Xeon E5-2697 v3 CPU (2.6 GHz) with 28 cores and 128 GB memory

Program language: C++

Software required: Geographic Data Abstraction Library (GDAL). Source code available at the link: https://gdal.org/download.html. For Unix systems it is also available at the repositories with the name libgdal-dev.

Program size: 153 KB

The source codes are available for downloading at the link: https://gitlab.kuleuven.be/u0123674/acamf

# References

Anand, S.K., Hooshyar, M., Porporato, A., 2020. Linear layout of multiple flow-direction networks for landscape-evolution simulations. Environmental Modelling & Software 133, 104804. URL: https://www.sciencedirect.com/science/article/pii/S1364815220305934, doi:https://doi.org/10.1016/j.envsoft.2020.104804.

B.-M. Vought, L., Pinay, G., Fuglsang, A., Ruffinoni, C., 1995. Structure and function of buffer strips from a water quality perspective in agricultural landscapes. Landscape and Urban Planning 31, 323–331. URL: https://www.sciencedirect.com/science/article/pii/016920469401057F, doi:https://doi.org/10.1016/0169-2046(94)01057-F.

Cisneros Espinosa, P.J., Gabriëls, D., Van Meirvenne, M., 1999. Mapping of soil erosion risk zones in a part of the watershed of rio Paute (Ecuador). URL: http://lib.ugent.be/catalog/rug01:000500575.

Conrad, O., 2003. Tool LS Factor. URL: https://saga-gis.sourceforge.io/saga_tool_doc/8.0.1/ta_hydrology_22.html. Accessed: 2022-01-11.

Costa-Cabral, M.C., Burges, S.J., 1994. Digital Elevation Model Networks (DEMON): A model of flow over hillslopes for computation of contributing and dispersal areas. Water Resources Research 30, 1681–1692. URL: https://agupubs.onlinelibrary.wiley.com/doi/abs/10.1029/93WR03512, doi:https://doi.org/10.1029/93WR03512.

Costin, A., 1980. Runoff and soil and nutrient losses from an improved pasture at ginninderra, southern tablelands, new south wales. Crop & Pasture Science 31, 533–546.

Deproost, P., Renders, D., Wauw, J., Ransbeeck, N., Verstraeten, G., 2018. Herkalibratie van WaTEM/SEDEM met het DHMV-II als hoogte- model omgevingvlaanderen.be Eindrapport. Technical Report. KU Leuven. Belgium.

Desmet, P., Govers, G., 1996. A GIS procedure for automatically calculating the USLE LS Factor on topographically complex landscape units. Journal of Soil and Water Conservation 51, 427–433.

Drzewiecki, W., Mularz, S.C., 2008. Simulation of water soil erosion effects on sediment delivery to dobczyce reservoir.

Estrella, R., 2015. Where to afforest? Single and multiple criteria evaluation methods for spatio-temporal decision support, with application to afforestation. Ph.D. thesis. Faculty of Bioscience Engineering, KU Leuven. URL: https://lirias.kuleuven.be/retrieve/345463.

Estrella, R., Cattrysse, D., Van Orshoven, J., 2014a. Comparison of three ideal point-based multi-criteria decision methods for afforestation planning. Forests 5, 3222–3240. URL: https://www.mdpi.com/1999-4907/5/12/3222, doi:10.3390/f5123222.

Estrella, R., Vanegas, P., Cattrysse, D., Van Orshoven, J., 2014b. Trading off accuracy and computational efficiency of an afforestation site location method for minimizing sediment yield in a river catchment, Rückemann, Claus-Peter. International Academy, Research, and Industry Association ( IARIA ). pp. 94–100. URL: https://lirias.kuleuven.be/retrieve/267799.

Fischer, D.T., Church, R.L., 2003. Clustering and Compactness in Reserve Site Selection: An Extension of the Biodiversity Management Area Selection Model. Forest Science 49, 555–565. URL: https://doi.org/10.1093/forestscience/49.4.555, doi:10.1093/forestscience/49.4.555.

Gabriels, K., Willems, P., Van orshoven, J., 2022. An iterative runoff propagation approach to identify priority locations for land cover change minimizing downstream river flood hazard. Landscape and Urban Planning 218, 104262. URL: https://www.sciencedirect.com/science/article/pii/S0169204621002255, doi:https://doi.org/10.1016/j.landurbplan.2021.104262.

Gersmehl, P., 1970. Spatial interaction. Journal of Geography 69, 522–530. URL: https://doi.org/10.1080/00221347008981861, doi:10.1080/00221347008981861.

Han, J., Kamber, M., Pei, J., 2012. Data Preprocessing, in: Han, J., Kamber, M., Pei, J. (Eds.), Data Mining. Third Edition ed.. Morgan Kaufmann, Boston. The Morgan Kaufmann Series in Data Management Systems, pp. 83–124. URL: https://www.sciencedirect.com/science/article/pii/B9780123814791000034, doi:https://doi.org/10.1016/B978-0-12-381479-1.00003-4.

Heil, G., Muys, B., Hansen, K., 2007. Environmental Effects of Afforestation in North-Western Europe – From Field Observations to Decision Support. doi:10.1007/1-4020-4568-9.

Jiang, L., Tang, G., Liu, X., Song, X., Yang, J., Liu, K., 2013. Parallel contributing area calculation with granularity control on massive grid terrain datasets. Computers & Geosciences 60, 70–80. URL: https://www.sciencedirect.com/science/article/pii/S0098300413001921, doi:https://doi.org/10.1016/j.cageo.2013.07.003.

Kahn, A.B., 1962. Topological sorting of large networks. Commun. ACM 5, 558–562. URL: http://doi.acm.org/10.1145/368996.369025, doi:10.1145/368996.369025.

Nearing, M., Jetten, V., Baffaut, C., Cerdan, O., Couturier, A., Hernandez, M., Le Bissonnais, Y., Nichols, M., Nunes, J., Renschler, C., Souchère, V., van Oost, K., 2005. Modeling response of soil erosion and runoff to changes in precipitation and cover. CATENA 61, 131–154. URL: https://www.sciencedirect.com/science/article/pii/S0341816205000512, doi:https://doi.org/10.1016/j.catena.2005.03.007. Soil Erosion under Climate Change: Rates, Implications and Feedbacks.

O'Callaghan, J.F., Mark, D.M., 1984. The extraction of drainage networks from Digital Elevation Data. Computer Vision, Graphics, and Image Processing 28, 323 – 344. URL: http://www.sciencedirect.com/science/article/pii/S0734189X84800110, doi:https://doi.org/10.1016/S0734-189X(84)80011-0.

Piégay, H., Walling, D., Landon, N., He, Q., Liébault, F., Petiot, R., 2013. Contemporary changes in sediment yield in an alpine montane basin due to afforestation (the upper-drôme in France). Catena , 183–212doi:10.1016/S0341-8162(03)00118-8.

Qin, C.Z., Zhan, L., 2012. Parallelizing flow-accumulation calculations on graphics processing units—from iterative dem preprocessing algorithm to recursive multiple-flow-direction algorithm. Computers & Geosciences 43, 7–16. URL: https://www.sciencedirect.com/science/article/pii/S0098300412000787, doi:https://doi.org/10.1016/j.cageo.2012.02.022.

Quinn, P., Beven, K., Chevallier, P., Planchon, O., 1991. The prediction of hillslope flow paths for distributed hydrological modelling using digital terrain models. Hydrological Processes 5, 59–79. URL: https://onlinelibrary.wiley.com/doi/abs/10.1002/hyp.3360050106, doi:10.1002/hyp.3360050106.

Randolph, J., 2012. Environmental Land Use Planning and Management. Island Press. URL: https://books.google.be/books?id=Kf83YgEACAAJ.

Renard, K.G., Foster, G.R., Weesies, G.A., Porter, J.P., 1991. RUSLE: Revised Universal Soil Loss Equation. Journal of Soil and Water Conservation 46, 30–33. URL: https://www.jswconline.org/content/46/1/30.

Tarboton, D.G., 1997. A new method for the determination of flow directions and upslope areas in grid Digital Elevation Models. Water Resources Research 33, 309–319. URL: https://agupubs.onlinelibrary.wiley.com/doi/abs/10.1029/96WR03137, doi:https://doi.org/10.1029/96WR03137.

Vanegas, P., 2010. A spatially explicit approach to the site location problem in raster maps with application to afforestation. Ph.D. thesis. KU Leuven, Belgium.

Vanegas, P., Cattrysse, D., Orshoven, J., 2012. Allocating reforestation areas for sediment flow minimization: an integer programming formulation and a heuristic solution method. Optimization and Engineering 13, 247–269.

Vanegas, P., Cattrysse, D., Van Orshoven, J., 2009. Compactness and Flow Minimization Requirements in Reforestation Initiatives: An Integer Programming (IP) Formulation, in: Gervasi, O., Taniar, D., Murgante, B., Laganà, A., Mun, Y., Gavrilova, M.L. (Eds.), Computational Science and Its Applications – ICCSA 2009, Springer Berlin Heidelberg, Berlin, Heidelberg. pp. 132–147.

Vanegas, P., Cattrysse, D., Van Orshoven, J., 2011. A multiple criteria heuristic solution method for locating near to optimal contiguous and compact sites in raster maps, in: Geocomputation, sustainablity and environmental planning. 1st ed.. Springer. volume 348, pp. 35–56.

Vanegas, P., Cattrysse, D., Wijffels, A., Van Orshoven, J., 2014. Compactness and flow minimization requirements in reforestation initiatives: A heuristic solution method. Annals of Operations Research 219, 433–456.

Vanwalleghem, T., 2017. Soil Erosion and Conservation. John Wiley & Sons, Ltd. pp. 1–10. URL: https://onlinelibrary.wiley.com/doi/abs/10.1002/9781118786352.wbieg0381, doi:https://doi.org/10.1002/9781118786352.wbieg0381.

Wang, J., 2017. Economic Geography: Spatial Interaction. John Wiley & Sons, Ltd. pp. 1–4. URL: https://onlinelibrary.wiley.com/doi/abs/10.1002/9781118786352.wbieg0641, doi:https://doi.org/10.1002/9781118786352.wbieg0641.

Wijffels, A., Van Orshoven, J., 2009. Contribution of Afforestation to the Enhancement of Physical and Socio-Economic Land Performance in the Southern Andes of Ecuador: Assessment, Modelling and Planning Support. Technical Report. Vlaamse Interuniversitaire Raad (VLIR). Leuven, Belgium.

Williams, J.C., Revelle, C.S., 1997. Applying mathematical programming to reserve selection. Environmental Modeling & Assessment 2, 167–175.

Wilson, J., Gallant, J., 2000. Terrain Analysis: Principles and Applications. Earth sciences: Geography, Wiley. URL: https://books.google.be/books?id=1311_4-zvy4C.

Wischmeier, W.H., 1978. Predicting rainfall erosion losses : a guide to conservation planning , 58 p. : ill., maps ; 26 cm. —USDAURL: https://handle.nal.usda.gov/10113/CAT79706928.